\documentclass{article}
\usepackage[utf8]{inputenc}
\usepackage{amsmath}
\usepackage{amssymb}
\usepackage{epstopdf}

\usepackage{graphicx}
\usepackage{authblk}
\usepackage{natbib}
\usepackage{hyperref}
\usepackage{setspace}
\usepackage[a4paper, total={6in,9in}]{geometry}

\vspace{-40pt}
\title{Visualising the Evolution of English Covid-19 Cases with Topological Data Analysis Ball Mapper} 
\vspace{-30pt}
\author[1]{Pawel Dlotko}
\affil[1]{Mathematics Department, Swansea University, Bay Campus, Swansea, SA8 1EN, United Kingdom}
\vspace{-40pt}

\vspace{-30pt}
\author[2]{Simon Rudkin\thanks{Corresponding author. Tel: (0044)1792 606159, Email:s.t.rudkin@swansea.ac.uk }}
\affil[2]{School of Management, Swansea University, Bay Campus, Swansea, SA8 1EN, United Kingdom}
\vspace{-20pt}
\begin{document}
\maketitle

\begin{abstract}
 Understanding disease spread through data visualisation has concentrated on trends and maps. Whilst these are helpful, they neglect important multi-dimensional interactions between characteristics of communities. Using the Topological Data Analysis Ball Mapper algorithm we construct an abstract representation of NUTS3 level economic data, overlaying onto it the confirmed cases of Covid-19 in England. In so doing we may understand how the disease spreads on different socio-economical dimensions. It is observed that some areas of the characteristic space have quickly raced to the highest levels of infection, while others close by in the characteristic space, do not show large infection growth. Likewise, we see patterns emerging in very different areas that command more monitoring. A strong contribution for Topological Data Analysis, and the Ball Mapper algorithm especially, in comprehending dynamic epidemic data is signposted.
\end{abstract}

\maketitle

\section{Introduction}

As Covid-19 spreads through the global community, the bridge between economics and data science offers much to unlocking the information within the official statistics. Where summary statistics speak of trends, and maps of cases help get a visual handle on the spatial scale, Topological Data Analysis (TDA) after \cite{carlsson2009topology} and particularly the Ball Mapper (BM) algorithm of \cite{dlotko2019ball} can quickly highlight patterns within the characteristics of communities for policy to attend to. This short note, firstly, contributes a first look at how BM produces an abstract two dimensional representation of NUTS3 data and how, by doing so, we can see where in the characteristic space cases are particularly fast rising in terms of number of infections. Whilst regional characteristics are comparatively static, the changing nature of cases facilitates the envisioning of changing outcomes on the point cloud. A second contribution is thus to show how the relative levels of Covid-19 are changing within the space. Moreover we bring to the empirical economics literature an approach that recognises the outcomes we study do not manifest linearly through space as our models may present, nor are they organised neatly along axes. Interactions matter, location in the space is important, and visualising the true picture from our data is the only way to see and quantify what is truly there.

Intuition for the BM plot is as straightforward as the wealth of information we would gain from seeing a two-dimensional scatter plot of two characteristics. The BM technique allows us to visualize multiple dimensional data sets represented in the form of point clouds. Having this helps to understand where the considered activity, in this case the infection spread, is most prevalent. As an observation has a coordinate on a scatter plot so it is a point in this multi-dimensional cloud. 
Where regions in that cloud have high incidences of Covid-19 so attention must focus on the combination of each characteristic that is associated therewith; just as we would look at whether high incidences were top left, top right, etc. in a scatter plot. The dynamics are thus quickly viewed and trends therein exposited.   

\cite{adda2016economic} speaks much to the present situation with Covid-19, placing the spread of epidemics firmly within a world of increased transportation and trade. The high frequency use of influenza data in \cite{adda2016economic} aids economists in understanding how economic conditions link to disease spread, engaging a discussion with epidemiology. This contribution to the literature owes much to the pioneering work in combining high, and low, frequency data to marry annually updated regional characteristics with the daily information on Covid-19 cases. Obtaining data at a sufficiently micro level for study in the UK presents its own challenges, met in this paper by the use of NUTS3 level data\footnote{NUTS is the acronym for the Nomenclature of territorial units for statistics. It is the hierarchical system for dividing up the geographic regions of the European Union and the United Kingdom. NUTS3 units are the smallest level of aggregation available from the European Union's Eurostat database.}. A strong focus in the discussion of economics and disease spread is placed on employment, following works on influenza by \cite{markowitz2019effects} and \cite{eichenbaum2020macroeconomics}. Likewise there are considerations of income, working patterns and population age that can help understand spread \citep{bockerman2009economic, barnay2016health}. Population density is a natural key to contamination rates also and must be included \citep{jones2008global}. Familiarity with what propagates the spread of common flu viruses gives strong guidance to what may then drive the spread of Covid-19. This paper maps the characteristics space for UK NUTS3 regions to the extent of available data.

Early work to link Covid-19 to economic conditions in China has sought data science to explain which factors are important to spread \citep{qiu2020impacts}, but such Machine Learning approaches are subject to overfitting criticisms. Unlike the ``black box'' nature of non-linear and deep-learner models, the BM approach here is simply a representation of the data that the user interprets. There is also much to remind of other visualisation methods, such as t-SNE \citep{maaten2008visualizing} as endorsed recently in medical analysis by \cite{linderman2019fast}, and the original TDA mapper algorithm of \cite{singh2007topological}. Unlike these approaches BM does not require the construction of a full distance matrix between points as t-SNE does, and is more stable than the original mapper as BM has just one parameter and no functional choices. Consequently the insight gained into economics and the spread of disease from this paper have full transparency. In what follows we take the learning from past studies of disease spread, the methodological advancements of TDA and BM and explore what we may learn from the evolving English picture. 

\section{Methodology}

BM is a means of representing complex and high dimensional data sets as abstract graphs. After covering them with a collection of balls, it creates a visualisation of the considered multidimensional phenomenon as an abstract two-dimensional form. BM graphs are constructed by covering the considered data set with a collection of overlapping balls of a radius $\epsilon$. 
Central points of each ball are referred to as landmarks, the number of landmarks being a function of the density of the data and the choice of $\epsilon$. Where there are points in the intersection of two balls those balls are connected by an edge on the graph. Numbers of points covered by a ball guide its size in the graph. Consequently we see the density of data, connectivity and relationships between observations in multi-dimensional space.

Overlaying the BM graph with colour facilitates the consideration of an outcome of interest, here the number of cases of Covid-19. By default the colour of a ball picks up the average outcome for the points within the ball, but if desired the colouring function may be altered by the user. As implemented in the Ball Mapper R package~\cite{dlotko2019R} a colour bar is added to the plot whereby the lowest values are red and the highest are purple. Labels on the bar direct on the specific meanings of colour for that plot. Seeing pairs of connected balls with very different colouration immediately attracts attention. Recalling that the balls link to observations in the underlying dataset we may quickly merge back to the data set using the point ID to link that data with the ball labels. Once this is done it is possible to acquire a more detailed picture of what has been found, including non-continuous variables that had been necessarily kept out of the cloud formation. In this case the region name is a relevant example of something it is useful to match. Grouping by ball offers summary statistics and a chance to measure differentials in the euclidean space.



\section{Data}

Data on NUTS3 regions is taken from Eurostat, whilst data on earnings in the UK is collected from the Earnings and hours worked, place of work by local authority: ASHE Table 7 survey produced by the Office for National Statistics (ONS) in the UK. Cases of Covid-19 by local authority are reported by the UK government and are as collated for researchers on the Github site of Tom White\footnote{The URL for this site is  \href{https://github.com/tomwhite/covid-19-uk-data}{https://github.com/tomwhite/covid-19-uk-data} where you may find the code for the retrieval and processing of data from public facing UK government websites.}. Matching from local authority to NUTS3 is performed using the look up queries from the Office for National Statistics, supplemented by some manual work on cases where no match can be found. Because of the nature of reporting, the data used here only considers England.

\begin{table}
    \begin{center}
        \caption{Summary Statistics of NUTS3 Data}
        \label{tab:sumstats}
        \begin{tabular}{l c c c c}
            \hline
            Characteristic & Mean & s.d. & Min & Max \\
            \hline
            Population Density (Population per $KM^2$) & 2454 & 1337 & 15 & 4734\\
            Median Age & 40.52 & 4.742 & 29.90 & 50.40 \\
            Gross Domestic Product (\pounds 000's) & 15719 & 15067 & 2517 & 131476 \\
            Gross Value Added (\pounds 000's) & 13991 & 13411 & 2240 & 117025 \\
            Average Hours Worked & 33.37 & 0.854 & 31.00 & 35.50\\
            Average Annual Gross Income (\pounds 000's) & 17.03 & 3.501 & 12.25 & 35.92\\
            \hline
        \end{tabular}
    \end{center}
\raggedright
\footnotesize{Notes: Data on population density, median age, regional Gross Domestic Product (GDP) and regional Gross Value Added (GVA) is collected at the NUTS3 level and downloaded from Eurostat. Data on average hours worked and average annual gross income is taken from the Earnings and hours worked, place of work by local authority: ASHE Table 7 survey produced by the Office for National Statistics (ONS) in the UK. Case data is released by the UK Government via \href{www.gov.uk}{www.gov.uk} at the local authority level. Merging from local authority to NUTS3 is performed using the look up queries offered by the ONS and supplemented with manual matching. Where one local authority covers multiple NUTS3 codes cases, GDP and GVA are spread evenly amongst those codes. ($n=134$)}
\end{table}

Summary statistics for the variables used are provided in Table \ref{tab:sumstats}, with the six characteristics of the point cloud showing very different scales. To overcome the challenge of having one dominant axis, the axes of the dataset are normalised into the interval $[0,1]$ so that the maximal value of each column gets mapped to $1$, the minimal value gets mapped to $0$, and all the intermediate values are mapped inbetween. In this paper raw values of Gross Domestic Product (GDP) and Gross Value Added (GVA) are used as our aim is to distinguish tangibly between regions. Employing the log--scaling would create greater spread lower down the distribution, meaning more balls at the lower income level,  but would equally connect very different high income areas at the top of the scale. Average hours worked do not vary greatly but we do see a large regional income disparity; such is well studied for the United Kingdom. Because this is economic data, appeal is made to the geographic imbalance of the United Kingdom when considering spatial extents; weighting of GDP and wages towards London means they serve to capture much of the geography of the spread. Likewise on a micro scale the movement away from regional economic centres like Manchester, Bristol or Newcastle conveys a similar spatial extent.

\begin{table}
    \begin{center}
        \caption{Daily Covid-19 Cases}
        \label{tab:covidstats}
        \begin{tabular}{l c c c c}
            \hline
            Date & Mean & s.d. & Min & Max \\
        \hline
        14th March &2.341&2.828&0&16\\
        15th March &2.705&3.146&0&18\\
        16th March &2.705&3.146&0&18\\
        17th March &4.053&4.608&0&26\\
        18th March &5.788&6.799&0&37\\
        19th March &7.386&8.94&0&59\\
        20th March &8.402&9.866&0&64\\
        21st March &10.77&12.90&0&76\\
        22nd March &15.64&17.31&0&92\\
        24th March &19.27&21.65&0&121\\
        25th March &30.27&33.76&0&194\\
        26th March &35.02&37.46&1&206\\
        27th March &40.92&42.37&1&240\\
        28th March &50.05&50.12&1&278\\
        29th March &58.31&57.24&2&323\\
        30th March &70.64&69.34&4&397\\
        31st March &89.18&81.82&5&448\\
        1st April &105.0&94.52&5&507\\
        2nd April &120.2&106.8&9&583\\
        3rd April &135.6&116.9&11&616\\
        4th April &153.0&127.1&11&696\\
        5th April &180.0&147.3&14&759\\
        6th April &206.1&165.0&17&862\\
        7th April &232.2&183.4&19&984\\
        8th April &253.0&194.0&25&1138\\
        9th April &288.3&218.1&31&1246\\
        10th April &312.6&228.6&32&1287\\
        11th April &334.5&240.7&36&1372\\
        12th April &369.4&257.2&38&1462\\
        13th April &396.4&272.9&41&1524\\
        14th April &427.0&285.7&41&1604\\
        15th April &454.6&299.4&45&1712\\
        16th April &481.3&311.2&53&1774\\
        17th April &502.7&321.6&56&1853\\
        \hline
        \end{tabular}
    \end{center}
\raggedright
\footnotesize{Numbers represent the reported number of Covid-19 cases in total for a given NUTS3 code within England. In each case the number reported is the total number over all days before and including that date. Statistics are as released by the UK Government ( \href{www.gov.uk}{www.gov.uk} ) at the local authority level and are translated to the NUTS3 level for this analysis. Collation of the daily figures is via Tom White on Github \href{https://github.com/tomwhite/covid-19-uk-data}{https://github.com/tomwhite/covid-19-uk-data}. A change in reporting time means that 23rd of March is omitted.}
\end{table}

Covid-19 case data is updated daily and may be neatly summarised in Table \ref{tab:covidstats}. The rise in cases is clear, but equally the growth of the standard deviation reminds that the disparities across regions are large. These numbers are the total number of cases reported up to, and including, that date. At this time the trend continues upwards and the need to understand more of the spatial economics behind these rises is clear.

\section{Results}

In the first instance we are concerned with the evolving picture of cases in England. The BM graph is constructed using the six characteristics of the NUTS3 regions shown in Table \ref{tab:sumstats}, recalling that these are normalised onto the interval $[0,1]$ owing to the varying scales of the different regional economic measures. In choosing the radius for the balls we considered carefully the trade-off between noise in the colouration and being able to chart neatly the dynamics of the cases. Colouration of the balls is a three step process. Firstly the average number of cases on a given day within the ball is calculated. This number is then divided by the sum of cases for all balls at that day. That gives a proportion of the cases within the considered ball to all cases at the given day. Note that this is done in this way because points appear in multiple balls where they sit in the intersection. The resulting variable is then passed back to the BM graph which is subsequently plotted. Altering the colouration in this way reminds that the total number of cases is increasing every day and ensures the balls only show the relative position. The alternative of calculating the new cases is liable to be volatile, high infection areas may report low cases on one day and appear lower risk for example. Hence using our adjustment allows the continued charting of dynamics.

\begin{figure}
   \begin{center}
       \caption{Evolution of Cumulative Case Numbers }
       \label{fig:days}
       \begin{tabular}{c c c}
            \multicolumn{3}{c}{\includegraphics[width=12cm]{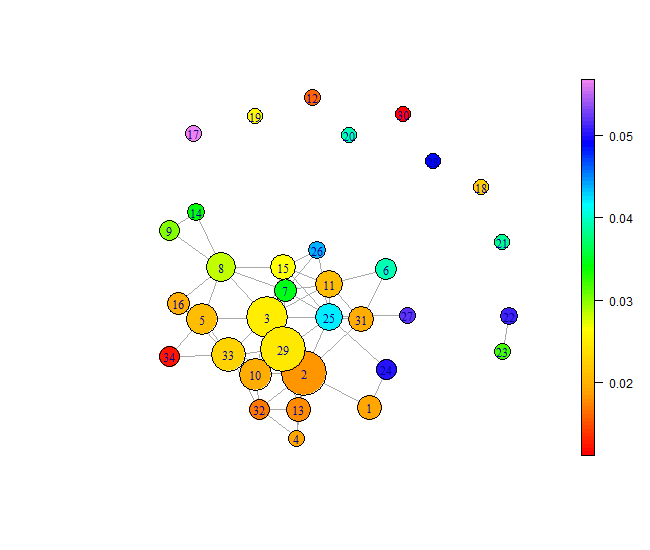}}\\
            \multicolumn{3}{c}{(a) 17th April 2020} \\
            \includegraphics[width=4cm]{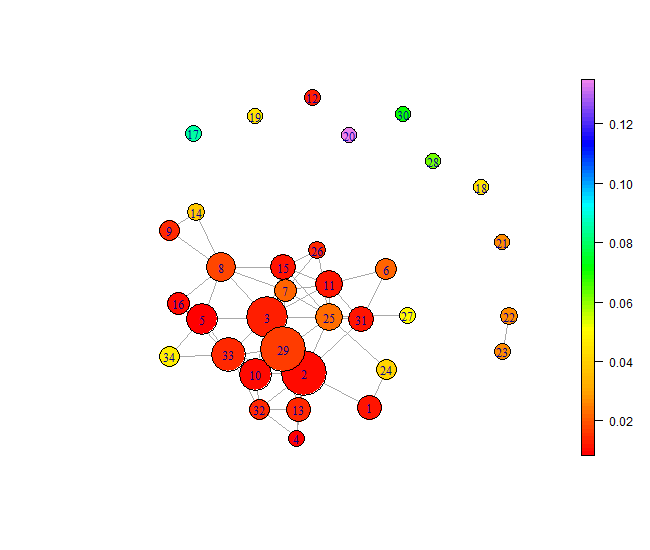} &
            \includegraphics[width=4cm]{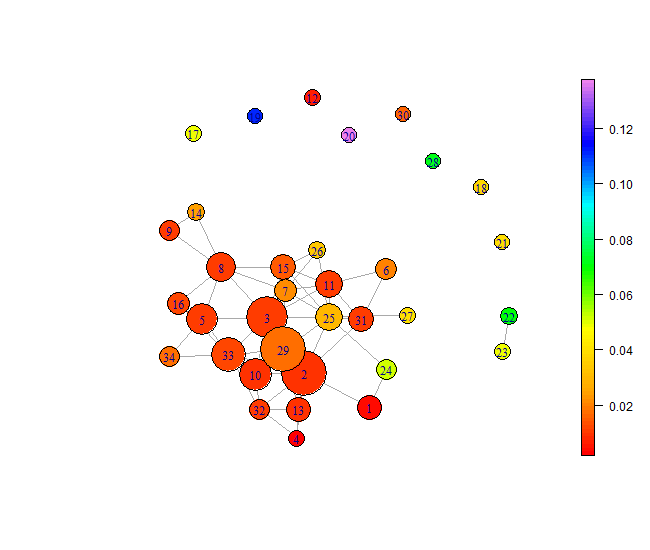} & \includegraphics[width=4cm]{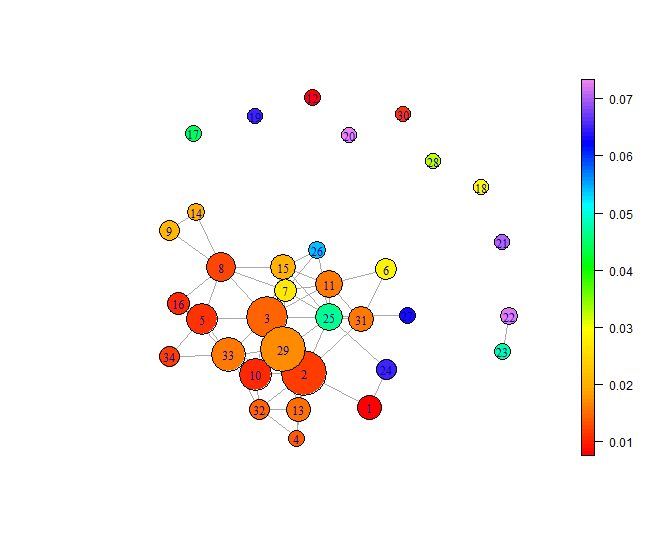}\\
            
            (b) 14th March 2020 & (c) 21st March 2020  
             &  (d) 28th March 2020 \\
             \includegraphics[width=4cm]{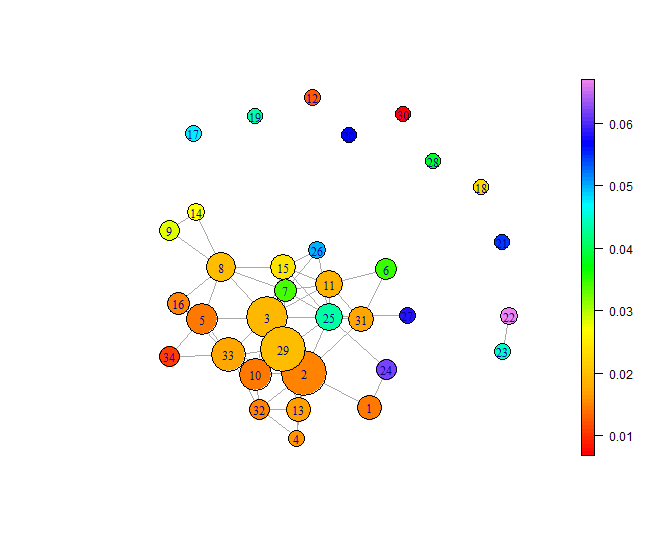} &
             \includegraphics[width=4cm]{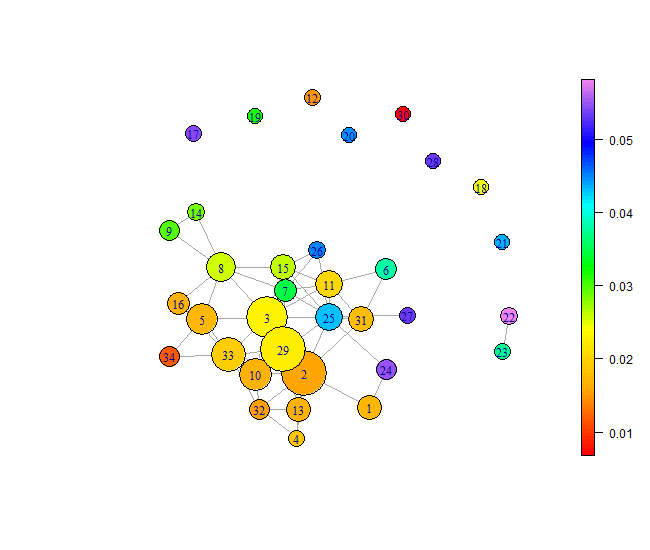} & \includegraphics[width=4cm]{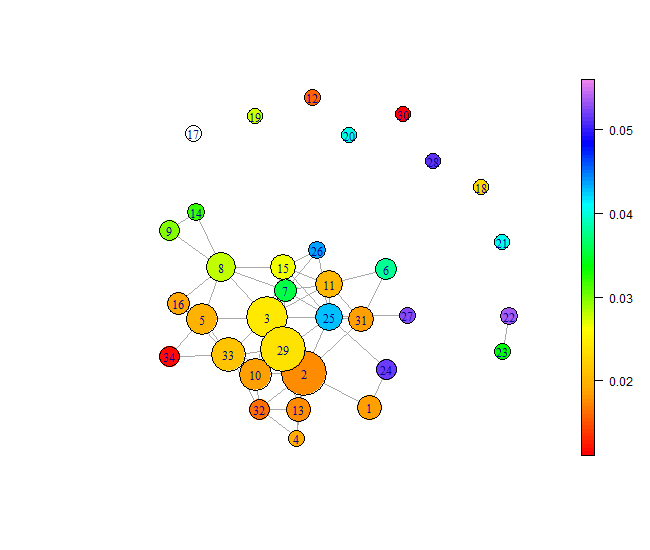}\\
             (e) 3rd April 2020 & (f) 10th April 2020  
             &  (g) 14th April 2020 \\
            
       \end{tabular}
   \end{center}
 \raggedright{Notes: Axes used to construct the BM graph are, Population Density, Median Age, Gross Domestic Product (GDP) and Gross Value-Added (GVA), sourced at the NUTS3 level from Eurostat, and Hours Worked and Gross Pay are from the UK Labour Force Study. Plots generated using \texttt{BallMapper}. Colouration is the proportion of overall English cases which are within each ball. Blues and purples correspond to the highest proportion of cases.}
\end{figure}

Figure \ref{fig:days} shows four plots which chart the changing picture of Covid-19 cases in England. These BM graphs are drawn with six axes and demonstrate a broad connectivity between most regions of England. However there are a series of balls which do not connect at this value of $\epsilon$. These include those areas with particularly low population density, coloured red as they have a low number of cases, and those with particularly high earnings such as the combined region of Hammersmith, Fulham, Kensington and Chelsea, coloured purple. Indeed Ball 20, which represents this NUTS3 area is the highest in three of the four plots. Panels (b) to (d) show a broad consistency in the relative proportion of cases in each ball. By the time we reach 3rd April, panel (a) shows the dominance of ball 20 starts to break, with more blues emerging and a general move into the orange across the plot. What we see in panel (d) is a much more even distribution of the cases of the space. Since information about the NUTS3 areas within each ball is known we can quickly check the implications of the transmission. Whilst these new hotspots have similar population densities, median ages, GDP, GVA and hours worked there is notable differential in the average pay. It is the large differential in pay that leaves ball 20 disconnected from the space. From a socio-economic standpoint this transition into the hotspots being in areas of lower wage will be noteworthy in informing the policy response.

Because the colouring of each ball shows the relative number of cases within this ball to the global number of cases in all balls for the day, there is not a noticeable change in colour across the first three panels of Figure \ref{fig:days}. Rather, interest instead lies in the way that different subsets of the space change their colouration. Moving from (b) to (f) we see many balls catching up with ball 20, this is particularly true of the balls to the right of the connected shape. In this setting Ball 1 is interesting as it remains red whilst it's neighbour, ball 24, becomes progressively bluer. In early April the brightening of the connected component into orange has coincided with the replacement of ball 20 as the highest number of cases by ball 22. Panels (e) to (g) show the gradual lightening of the London balls, especially ball 20 which is green in the most recent plot of panel (a). The ball 1 to ball 24 contrast remains stark although recent plots are showing a closing of the gap.

\begin{figure}
   \begin{center}
       \caption{Axes Variables}
       \label{fig:av}
       \begin{tabular}{c c c}
            \includegraphics[width=4.6cm]{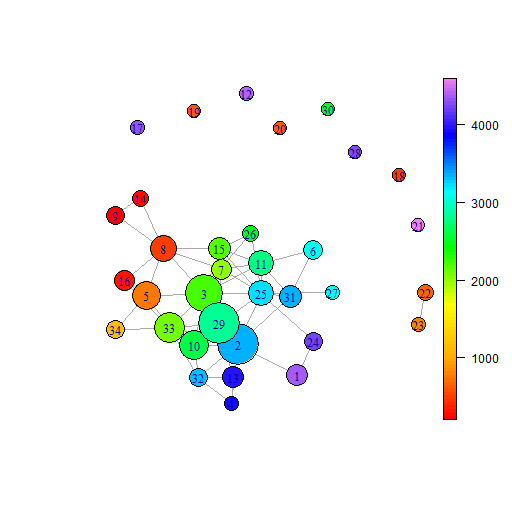} &
            \includegraphics[width=4.6cm]{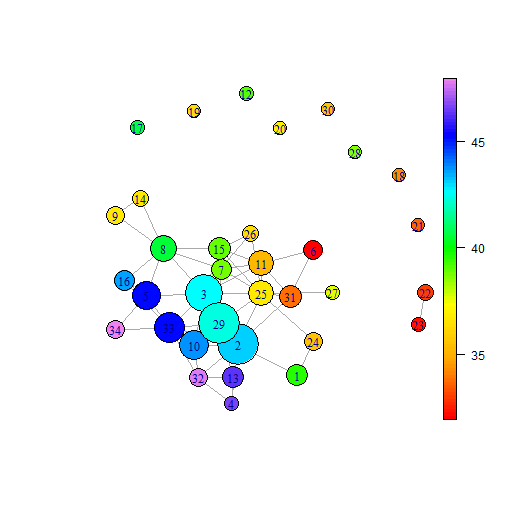} &
            \includegraphics[width=4.6cm]{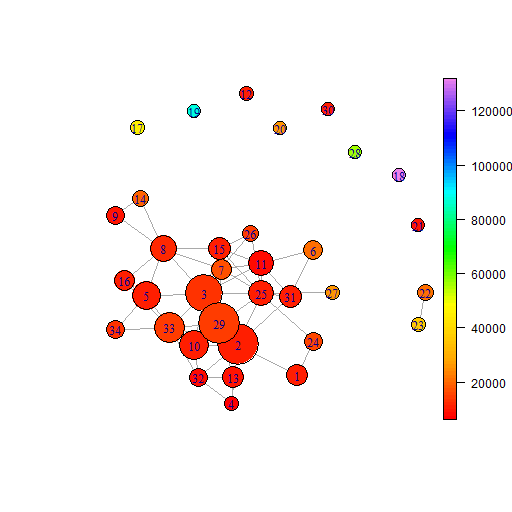} \\
            (a) Population Density & (b) Median Age & (c) GDP \\
            \includegraphics[width=4.6cm]{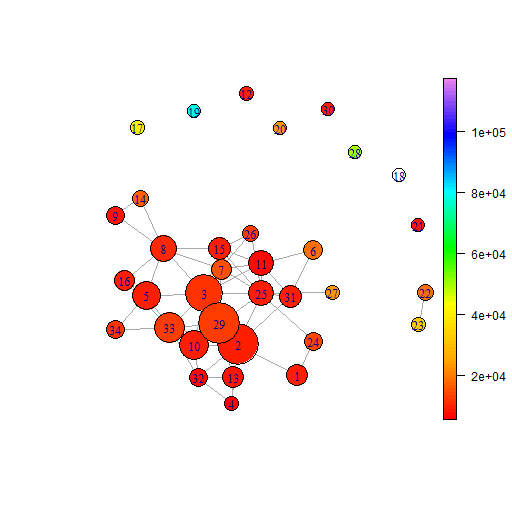} &
            \includegraphics[width=4.6cm]{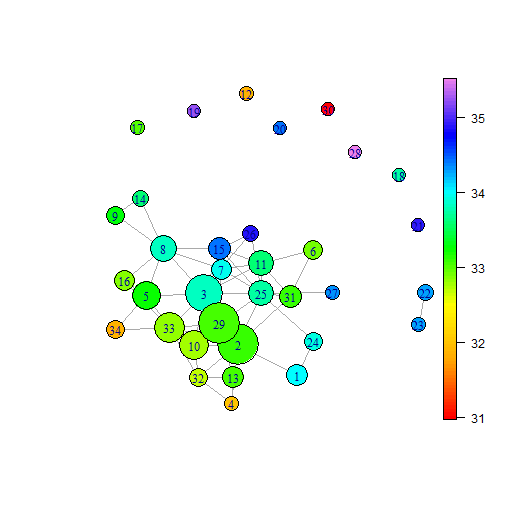} &
            \includegraphics[width=4.6cm]{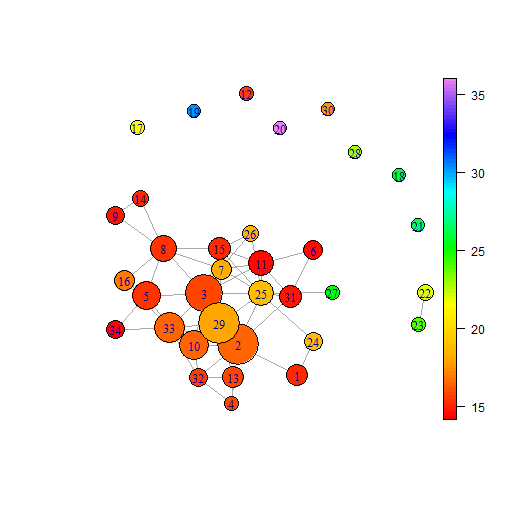} \\
            (d) GVA & (e) Hours Worked & (f) Gross Pay \\
            
       \end{tabular}
   \end{center}
 \raggedright{Notes: Population Density, Median Age, Gross Domestic Product (GDP) and Gross Value-Added (GVA) are sourced at the NUTS3 level from Eurostat. Hours Worked and Gross Pay are from the UK Labour Force Study. Due to reporting aggregation only England is considered. Plots generated using \texttt{BallMapper}.}
\end{figure}

To understand the plots further we may colour by the values of each axis instead of the number of Covid-19 cases. Figure \ref{fig:av} offers such a plot with one panel for each of the six axes. Panel (a) shows that it is the balls to the lower right of the main shape that have the highest population density, but the colouration of these is very mixed, the right-most set are where the numbers of cases increase notably but the left set do not see any increase at all. This suggests further confounding factors at play. Panel (b) may offer an explanation as those regions with the higher median age do correlate neatly with the ones that have the lowest reported number of Covid-19 cases. GDP and GVA, panels (c) and (d) are similar across the connected shapes as we know it is these two variables that drive the separation of the outlier balls. Likewise there is some similarity in gross pay, panel (f), though there we do witness differentiation between higher rates to the lower right and low rates at the top left. Hours worked do vary greatly with some correlation suggested there with Covid-19 cases.

BM graphs offer informative overviews of data, but their value comes in identifying the driving forces of those patterns. In the results we saw how there can be very different behaviour in balls that are similar enough to be connected. The most obvious case being balls 24 and 1. The former is predominantly districts of Outer London, whilst the latter is smaller cities such as Peterborough, Swindon, Hartlepool and the Thames Gateway. Connection between the two balls comes from Milton Keynes lying at the intersection. With the exception of Hartlepool, these are places which would see strong commuter links to London and yet the contrast with the balls that contain London is stark. Similar movement towards blue is seen in ball 27, which also contains London boroughs. 

There is a second interest at the other end of the connected shape where we see balls 9 and 14 become more yellow. This is not taking them to the same scale as the highest set, but it is evidence that there are non-uniform changes in other parts of the shape. These balls are the industrial north, including North-East Manchester, Blackburn and Darwen, Bradford and Sheffield. They are not major urban centres, and their overall population density is relatively low compared to other build up areas. Whether this transition is indicative of a growing cluster is unknown, but such changes are something which authorities may wish to monitor as more data comes in. The cities of Manchester, Liverpool, Nottingham, Birmingham and Bristol sit in ball 7, this lies between the major growth balls and balls 9 and 14. A suggestion is made that these serve as a conduit to their satellites, but ball 8 that has characteristics between the two, does not show such change in early April. As time has gone on we see distinct increases in the relative proportions through ball 8 into balls 9 and 14. The latter remain ahead of their neighbours. Our dynamic plots help greatly in charting change.

\begin{figure}
    \begin{center}
        \caption{Evolution of Average Cases per Ball}
        \label{fig:graph}
        \begin{tabular}{c c}
             \includegraphics[width=7cm]{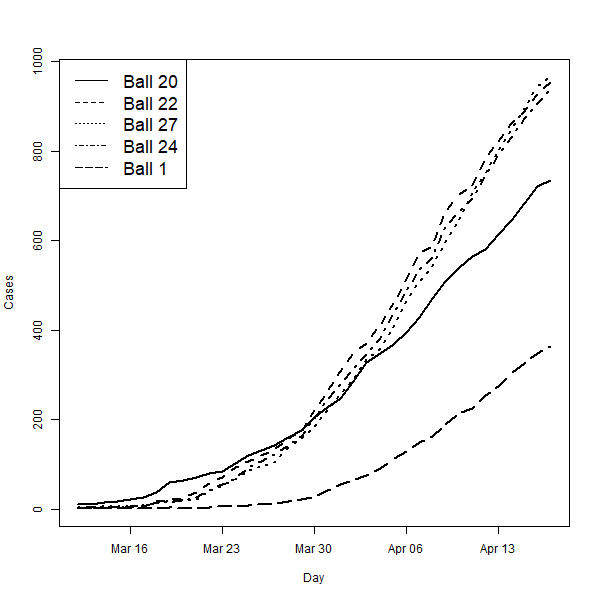} & \includegraphics[width=7cm]{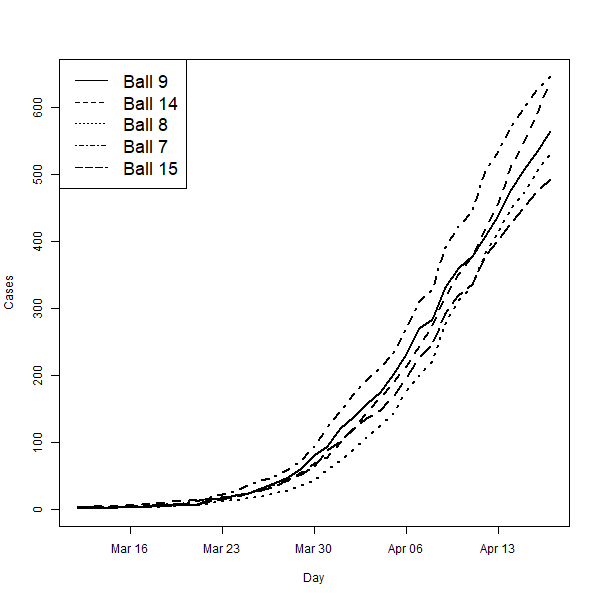}\\
             (a) London Boroughs and Commuters & (b) Northern Regions \\ 
        \end{tabular}
    \end{center}
\raggedright
\footnotesize{Notes: For panel (a) London Boroughs and Commuters is used to refer to the discussion around the highest proportion of cases. Panel (b) is labelled as Northern Regions to recognise the association of the story with towns linked to Manchester and Leeds. In each case the plot shows the averaged total number of cases reported up to the given date for each ball. Ball numbers refer to those used throughout this paper and are as featured in Figure \ref{fig:av}.}
\end{figure}

To chart the evolution of these stories Figure \ref{fig:graph} plots the number of cases in each of the main balls of interest. Panel (a) shows ball 20, the original maximum, as a solid line with the new maximum, ball 22, added as a dashed line. Ball 27, another in the connected shape that takes on dark blue hues is represented as a dotted line. Balls 24 and 1 are shown with dot-dash and long-dash lines respectively. The legend in the top left of the plot details. We see how progressively balls 22, then 24, then 27 have recorded more cases than ball 20. The London boroughs are thus gaining more cases than the original hotspot of Hammersmith and Fulham \& Kensington and Chelsea. A strong increase in recent days in Ball 1 is picked up towards the lower panel, interestingly occurring a week after the lockdown became effective on March 23rd. Through April the clear flattening of the curve in ball 20 can be seen but the other curves show limited evidence of flattening.

Panel (b) reflects on the two balls to the top left of the plot where the proportion of cases rose quickest, balls 9 and 14. These link through the characteristic space to London via balls 7 and 8, the former including Manchester, Leeds and other cities that would attract commuters from the areas of 9 and 14. Region 7 contains suburbs of Manchester and serves as a link to 9 and 14. Of all the regions plotted it is ball 7 that consistently has the lowest number of cases.

As the Covid-19 pandemic grows, so these evolutions will continue to interest. None of our plots suggest any flattening of the new cases curve anywhere except ball 20. Indeed the suggestion that other balls are starting on their own curves will be of concern to those watching the spread of Covid-19. Exploring these developments across the characteristic space will be a rich seem of work as the data develops.

\section{Summary}

Using a highly aggregated dataset of NUTS3 level economic indicators, this paper has charted the spread of the Covid-19 virus across England. The BM algorithm has been employed to represent the economic characteristic space. We have done so across six key axes, giving a mapping of the importance of interactions of these characteristics in determining outbreak extents. Developments in Outer London, and particularly the contrast with relatively similar areas which serve London as commuter towns, will be of interest in modelling the spread. Likewise, the growing intensity seen in the North of England, in the industrial heartlands of Lancashire and Yorkshire, will also be something for further thought for UK policy-makers. Development of the dataset, fine graining the information to better represent target populations and understanding more economic variables, will all help enhance the modelling. A further omission is the spatial dimension, there is no measure here to capture the physical geography; rather we simply appeal to the idea that many of our measures are driven by the distance to London because of the economic imbalance of the United Kingdom. Notwithstanding these concerns, the goal of this paper has been a demonstration of what BM can do and inspire the growth of its use. Our dynamic evaluation is constantly updating, and as more is learned so the narrative takes shape. Extending to other countries where localised reporting of cases is available offers a rich seam for impact generating exploration. Important steps are taken in both opening new perspectives to statistical studies in economics, but also in appreciating the spread of disease across socio-economic space.

\bibliography{covid}
\bibliographystyle{apalike}
\end{document}